# Voltage-Controlled Rotation of Magnetic Anisotropy in the Ni$_{90}$Fe$_{10}$/BaTiO$_3$(001) Heterostructure

A. Begué[1,2], M. W. Khaliq[3], N. Cotón[1], I. Lorenzo-Feijoo[1], M. A. Niño[3], M. Foerster[3], R. Ranchal[1,4,*]

[1]. Dpto. Física de Materiales, Fac. CC. Físicas. Universidad Complutense de Madrid, Plaza de las Ciencias 1, Madrid 28040, Spain

[2]. Fac. Ciencias, Universidad de Zaragoza, C. de Menéndez Pelayo 24, Zaragoza 50009, Spain

[3]. ALBA Synchrotron Light Facility, Carrer de la llum 2-26, Cerdanyola del Vallès 08290, Spain

[4]. Instituto de Magnetismo Aplicado, UCM-ADIF-CSIC, Las Rozas 28232, Spain

[*] Corresponding author: rociran@ucm.es

**Abstract**

In this work, we demonstrate the voltage control of magnetic anisotropy in a strain-mediated Ni$_{90}$Fe$_{10}$/BaTiO$_3$(001) heterostructure. In the pristine state of the heterostructure, the Magneto-Optical Kerr Effect measurements show a transcritical hysteresis loop for the Ni$_{90}$Fe$_{10}$ film, indicating a weak perpendicular anisotropy. This was further confirmed by X-ray Magnetic Circular Dichroism - Photoemission Electron Microscopy, revealing stripe domains in this film. X-ray diffraction analysis of the BaTiO$_3$ substrate under varying electric fields was used to analyze the orientation of ferroelectric domains. These results indicated that BaTiO$_3$ exhibits two distinct states depending on the applied electric field: one with domains aligned with the electric field and another with random domain orientation when the field is removed. After substrate poling, the Ni$_{90}$Fe$_{10}$ layer switches from weak perpendicular anisotropy to an in-plane uniaxial magnetic anisotropy, with the in-plane direction of anisotropy being controllable by 90° through an electric field. This effect is due to an efficient strain transfer from BaTiO$_3$ to the Ni$_{90}$Fe$_{10}$ lattice, induced by ferroelectric polarization, as shown by XRD. Remarkably, this rotation of the magnetic anisotropy leads to an enhanced converse magnetoelectric coupling value of 1.43 μs/m, surpassing previously reported values for other BaTiO$_3$-based heterostructures by an order of magnitude. These results emphasize the potential of Ni$_{90}$Fe$_{10}$ alloys for next-generation magnetoelectric devices.

**Introduction**

The search for increased efficiency in technological processes has driven research into innovative methods for material manipulation, particularly in the area of magnetization control [1, 2, 3, 4]. One promising approach is the field of voltage control of magnetic anisotropy (VCMA), which has garnered attention for its capability to control the magnetization of magnetic materials using voltage. The converse magnetoelectric coupling, $\alpha_E = \mu_0 \frac{\Delta M}{\Delta E}$, quantifies the relationship between the applied electric field, $\Delta E$, and the resulting changes in magnetization, $\Delta M$ [5, 6, 7]. This parameter is a key indicator of how strongly the electric field and magnetization are coupled.

Several strategies have been explored to achieve VCMA, one of which involves single-phase multiferroics. These materials exhibit several ferroic phases, *i.e.* combining ferroelectricity and ferromagnetism. Notable examples include TbMnO$_3$, with $\alpha_E = 10^{-5}$ µs/m [8], Ni$_3$B$_7$O$_{13}$I with 10$^{-3}$ µs/m [9], and Cr$_2$O$_3$ at 4.1×10$^{-6}$ µs/m [10]. However, these values are relatively low compared to those reported for hybrid multiferroics, which integrate two distinct materials with different ferroic phases. Examples such as BiFeO$_3$/BaTiO$_3$ with $\alpha_E$ = 0.2 µs/m [11], NFO/PZT with 0.8 µs/m [12], and FeRh/BaTiO$_3$ with 16 µs/m [13] demonstrate significantly higher coupling constants than those found in single-phase multiferroics. These values make hybrid multiferroics more suitable for practical applications, providing greater opportunities for their use in VCMA applications [14, 15].

Among the different hybrid multiferroics suitable for VCMA systems, our focus will be on strain-coupled systems. These systems present strong converse magnetoelectric coupling and exhibit low heat dissipation losses [16, 17]. They consist of a magnetostrictive layer coupled to a ferroelectric substrate. When voltage is applied to the substrate, it undergoes deformation, which is transmitted to the magnetostrictive layer. This deformation, due to the Villari effect, affects the magnetic anisotropy of the magnetostrictive layer. To achieve high coupling, it is desirable to use substrates with high piezoelectric values and magnetostrictive layers with high magnetostriction and weak magnetic anisotropy, facilitating control from the strain of the substrate [14, 15].

Ferroelectric substrates with high piezoelectric constants are commonly made from Pb-based materials, such as Pb[Zr$_x$Ti$_{1-x}$]O$_3$ (PZT) [18, 19] and Pb(Mg$_{1/3}$ Nb$_{2/3}$)O$_3$–PbTiO$_3$ (PMN-PT) [20, 21, 22, 23, 24]. However, there is growing concern about the use of Pb in materials, leading to a search for alternatives. Among these alternatives, BaTiO$_3$ (BTO), a well-known ferroelectric material with a perovskite-like structure, has emerged as a promising candidate, particularly for VCMA systems operating at room temperature. BTO generates strain through three distinct mechanisms: photostriction, phase change, and voltage induction. Photostriction, where the material changes dimensions upon exposure to light, holds promise for applications in light-driven actuators and sensors [25, 26, 27]. Phase change deformation is based on the lattice distortion that occurs during phase transitions, such as from tetragonal to cubic, driven by temperature changes [28, 29, 30]. Voltage induction, on the other hand, relies on the piezoelectric properties of BTO and is particularly relevant for VCMA heterostructures [31, 32, 33]. For example, Fe/BTO [31], FeGa/BTO [34, 33], and FeRh/BTO [13, 32] have demonstrated the potential of BTO for VCMA applications.

For the magnetostrictive layer, we explore alternatives to rare-earth materials like Terfenol-D, despite its magnetostrictive properties, due to its environmental impact [35]. Instead, we focus on Ni-Fe alloys, which are commonly found in the literature and are environmentally friendly. This study focuses on the Ni$_{90}$Fe$_{10}$ composition rather than pure Ni, Fe, or other alloys to balance magnetostriction and minimize out-of-plane (OOP) magnetic anisotropy for in-plane (IP) magnetization control. The magnetostriction of the alloy is $\lambda_{\text{Ni}_{90}\text{Fe}_{10}}$ = -22 ppm [36], compared to $\lambda_{Ni}$ = -33 ppm and $\lambda_{Fe}$ = -9.3 ppm [37]. Although Ni exhibits higher magnetostriction, electrodeposited thick films present OOP anisotropy [38], that makes it unsuitable for IP rotation, as it requires excessive strain. Our previous studies indicate that at ~11 at. % Fe content, the anisotropy transitions from OOP to IP [38]. Beyond this point, increasing Fe content weakens magnetostriction, making Ni$_{90}$Fe$_{10}$ the optimal composition for maximizing magnetostriction while maintaining favorable anisotropy. Additionally, we have demonstrated the effectiveness of electrodeposited Ni$_{90}$Fe$_{10}$ in VCMA systems, reporting a converse magnetoelectric coupling of 0.205 µs/m on BTO(011) under the piezoelectric regime [39]. Here, we investigate magnetic anisotropy control on the BTO(001) surface, focusing on its ferroelectric behavior. Moreover, the use of electrodeposition offers a cost-effective and scalable alternative to sputtering or evaporation, enabling the fabrication of thicker films with ease. While these methods have previously shown good results for depositing magnetostrictive materials on BTO substrates, our work demonstrates that electrodeposited films can achieve even better performance for strain-coupled systems, enabling efficient VCMA and making them suitable for transfer to industrial applications.

In this work, we first examined the response of the BTO substrate using X-ray diffraction (XRD) to analyze the ferroelectric domain configuration during active poling and at electrical remanence. Subsequently, we investigated the magnetic anisotropy of the NiFe layer in response to changes in the BTO substrate caused by poling, using magneto-optical Kerr effect (MOKE) measurements and Photo Emission Electron Microscopy with X-ray Magnetic Circular Dichroism (XMCD-PEEM) contrast. A value of $\alpha_E = 1.43$ µs/m for the heterostructure was achieved through the application of 0.4 MV/m, which revealed a high coupling between the layer and the substrate, enabling the manipulation of the IP uniaxial anisotropy of the NiFe layer.

**Experimental section**

The $Ni_{90}Fe_{10}$ layer was electrodeposited onto a BTO substrate (MSE Supplies), which had been prepared with a metallic bilayer of Ti (15 nm)/Au (50 nm) evaporated on both sides to make both surfaces of the BTO conductive. These bilayers are essential for the electrodeposition of the NiFe layer and also for the application of a perpendicular electric field to the heterostructure (as depicted in Fig. 1a). The $Ni_{90}Fe_{10}$ layer, with a thickness of 1.5 µm, was deposited under controlled conditions using a PalmSens EmStat3 system in a three-electrode cell configuration at room temperature. In this setup, the top Au surface of the BTO substrate served as the working electrode, a platinum mesh was used as the counter electrode, and an Ag/AgCl (3 M NaCl) electrode acted as the reference. The electrolyte solution, water-based with a pH of 2.4, consisted of 0.4 M $H_3BO_3$, 0.017 M saccharine, 0.7 M $NiSO_4$, and 0.02 M $FeSO_4$. A constant deposition voltage of -1.20 V was applied without stirring, specifically chosen to promote the formation of magnetic stripe domains in the NiFe [36, 38, 40].The film thickness was controlled by monitoring the charge at the working electrode throughout the process. Following deposition, the composition of the NiFe layer was characterized using energy dispersive X-ray spectroscopy, performed with a JEOL JSM 6400 instrument operating at an acceleration voltage of 15 kV.

XRD in reciprocal space configuration was employed to investigate the structural properties of BTO under various applied electric fields. The measurements were conducted using a D8 Bruker diffractometer, utilizing Cu K$\alpha_1$ ($\lambda$ = 1.54056 Å) and Cu K$\alpha_2$ ($\lambda$ = 1.54439 Å) wavelengths. The lattice parameters used to calculate the expected position of diffraction spots in reciprocal space are $a = b$ = 3.994 Å and $c$ = 4.038 Å [41]. The {103} reflection was selected for its simplicity in interpretation compared to other reflections, *i.e.* {002}. This reflection offers a single peak in $\omega$, allowing precise intensity centering and simultaneous observation of the multiferroic domain configuration.

We conducted MOKE measurements at room temperature using the longitudinal configuration with *p*-polarized light. The experimental setup, which was custom-built, includes a 650 nm wavelength laser with a power output of 5 mW operating at 10 kHz, two polarizers, a photodetector, and an electromagnet capable of generating a magnetic field of up to 200 mT. The signal from the photodetector is processed by a Stanford research SR830 lock-in amplifier to isolate the Kerr effect signal. The applied magnetic field strength is monitored using an FH-55 magnetometer, and the voltage is supplied through a voltage source to a custom sample holder.

Magnetic imaging was carried out using XMCD-PEEM at the CIRCE beamline of the ALBA Synchrotron [42]. The imaging was performed using low-energy secondary electrons (with kinetic energies around 1-2 V) at the Fe $L_3$ absorption edge, approximately 707 eV photon energy. Magnetic contrast was achieved by subtracting pixel-wise images taken with circularly polarized x-rays of opposite helicities (left and right handed circular polarization).

**Results and discussion**

First of all, the behavior of the BTO substrate under several poling fields was examined using XRD to investigate how ferroelectric domains align in response to the applied electric field. The electric field was cycled through +0.8 MV/m, +0 MV/m, -0.8 MV/m, and back to -0 MV/m. Fig. 1b presents the reflections obtained from reciprocal space measurements. The expected reflection positions (guidelines) for the tetragonal structure of BTO are based on lattice parameters $a = b$ = 3.994 Å and $c$ = 4.038 Å [41]. Under both polarities, ±0.8 MV/m bias, only the (103) reflection exhibits intensity. In contrast, at $E$ = 0 MV/m, multiple reflections, including (103), (301), and (310), are observed. The (103) reflection is related to the *c*-axis of BTO pointing OOP, while the (301) and (310) reflections are associated with the *c*-axis lying in the plane. Inset of Fig. 1b provides a schematic representation of the polarization of ferroelectric domains at two electric field states, $|E| = 0.8$ MV/m (top scheme) and $E = 0$ MV/m (bottom scheme). At $|E| = 0.8$ MV/m, the $c$-axis points OOP, as indicated by the main contribution to the (103) peak. However, at $E = 0$ MV/m, XRD measurements reveal the presence of ferroelectric domains with the $c$-axis oriented in multiple directions, suggesting a random distribution of domains. These findings indicate that BTO(001) substrate used in this work does not exhibit the typical characteristics of a standard BTO with this orientation. In fact, this first characterization highlights the importance of characterizing the ferroelectric substrate prior to the magnetoelectric analysis of the hybrid NiFe/BTO magnetoelectric heterostructure. When the electric field is removed, the disordered arrangement of the BTO ferroelectric domains does not allow the polarization remanence. These results show that BTO has two distinct states: one at null poling, where the domains are randomly oriented, and another with saturated poling, where the domains align with the applied electric field.

The NiFe layer was studied by XRD under various electric fields applied to the heterostructure to analyze how the strain from BTO affects the NiFe lattice. Due to the low XRD intensity of the polycrystalline NiFe layer, reciprocal space measurements were unable to provide information about the IP NiFe lattice parameters and their relationship with the BTO ferroelectric domains. However, in the $\theta - 2\theta$ configuration, we previously reported two reflections from our electrodeposited NiFe: (111) and (002) [38]. The NiFe (111) reflection appears at 44.41°, while the BTO (002) reflection is at 44.86°. Due to the high intensity and width of the BTO peak, we could not resolve the NiFe (111). However, the NiFe (002) reflection is separated from the substrate peaks, allowing for clear analysis. From $\theta - 2\theta$ measurements, applying Bragg's law ($n\lambda = 2d_{hkl}sin\theta$) and considering the cubic structure of the NiFe alloy ($a = d_{hkl}\sqrt{h^2 + k^2 + l^2}$), it is possible to achieve information about the OOP lattice parameter. Fig. 1c shows how the (002) NiFe reflection can be shifted by applying an electric field to the heterostructure. At $|E| = 0.8$ MV/m, the NiFe lattice parameter in the OOP direction is 3.548 Å, while at $E = 0$ MV/m, it is 3.541 Å. Therefore, when the electric field is applied, the lattice parameter corresponding to the OOP direction increases compared to the no voltage state (inset of Fig. 1c). This increase in the OOP lattice axis with the applied electric field is similar to the behavior observed in BTO, where the ferroelectric domains show the same trend, demonstrating that the strain from BTO is adequately transferred to the NiFe layer. As a result, we can control the NiFe strain state by electrically polarizing the ferroelectric substrate and, due to the NiFe magnetostriction, manipulate its magnetic anisotropy.

Following the structural characterization of the heterostructure, we proceed with the magnetic characterization. The magnetostrictive NiFe layer in its as-grown state was investigated using MOKE (Fig. 2a), revealing the characteristic transcritical hysteresis shape typically associated with stripe domain structures [43, 44]. Additionally, the XMCD-PEEM image shown in Fig. 2b confirms the presence of these stripe domains. This behavior is consistent with the expected formation of stripe domains in electrodeposited $Ni_{90}Fe_{10}$ alloys with thickness exceeding 200 nm [36]. The presence of stripe domains can be quantitatively calculated, taking into account the quality factor $Q$:

$$Q = \frac{2K_{PMA}}{\mu_0 M_s^2} \quad (1)$$

where $K_{PMA}$ is the perpendicular magnetic anisotropy (PMA) energy and $M_s$ is the saturation magnetization [45]. In materials characterized by moderate or low PMA ($Q < 1$), stripe domains develop when the film thickness exceeds a theoretical critical value $t_{cr}$:

$$t_{cr} = 2\pi \sqrt{\frac{A_{ex}}{K_{PMA}}} \quad (2)$$

where $A_{ex}$ is the exchange stiffness. If $Q > 0.1$, stripe domains are wider than the layer thickness, whereas for $Q < 0.1$ they exhibit a periodicity equal to the layer thickness [45]. The $K_{PMA}$ for our samples can be obtained from the IP MOKE loop (Fig. 2a) as:

$$K_{PMA} = \frac{\mu_0 H_K M_s}{2} \quad (3)$$

Where $\mu_0 H_K = 20$ mT is the anisotropy field extracted from experimental measurements, and $M_s = 5.9 \times 10^5$ A/m for $Ni_{90}Fe_{10}$ [46]. A value of $Q = 0.027$ is obtained, consistent with the presence of a low PMA in the as-grown NiFe layer.

Before investigating the magnetoelectric coupling in the NiFe/BTO heterostructure, it is necessary to define the IP directions used in the MOKE analysis. Fig. 3a illustrates the $\phi$ IP directions, where $\phi = 0º$ ($\phi = 90º$) is parallel to what we call in the following the BTO[010] (BTO[100]) axis. This directional scheme is established within the saturating poling framework, corresponding to the left scheme of Fig. 1b. To evaluate the magnetoelectric coupling between NiFe and BTO, we applied an electric field perpendicular to the heterostructure as depicted in Fig. 1a to observe the resulting magnetic changes using MOKE. Fig. 3b presents the hysteresis loops of the NiFe layer at $\phi = 0º$, comparing the as-grown state with two states achieved after poling: one without poling (0 MV/m) and another with applied poling (-0.4 MV/m). The 0 MV/m state (red curve) shows the hysteresis loop after poling the heterostructure and returning to electrical remanence. In this state, the transcritical shape of the as-grown loop disappears, indicating a transition in magnetic anisotropy direction from OOP to IP. Moreover, it indicates that a uniaxial anisotropy arises parallel to the substrate directions (see Fig. 3b and the inset at 0 MV/m), with its easy axis along $\phi = 0º$ and its hard axis along $\phi = 90º$. When a bias is applied, the axis for $\phi = 0º$ changes from easy to hard (Fig. 3b), whereas for $\phi = 90º$, it changes from hard to easy (Fig. 3b inset). From these values, we can estimate the strain transferred from the substrate to the film, assuming that magnetoelastic anisotropy is the sole cause of the change in uniaxial magnetic anisotropy. The minimum strain needed to rotate the uniaxial anisotropy is given by $K_u = K_{me}$, where $K_u$ is the uniaxial anisotropy and $K_{me}$ is the magnetoelastic anisotropy. Using $K_u = \frac{1}{2}\mu_0 H_a M_s$ and $K_{me} = \frac{3}{2}\lambda \varepsilon Y$, with $\mu_0 H_a$ = 77 mT (from Fig. 3b), $M_s$ = 0.65 MA/m [46], and $Y$ = 216 GPa [47], we estimate that the strain ε transferred from the BTO to the NiFe layer is -0.35%. By considering the Poisson's ratio of Ni$_{90}$Fe$_{10}$, which is $\nu \sim 0.22$ [47], we can estimate that to achieve a rotation of the uniaxial anisotropy, an OOP strain of $-\nu \cdot \varepsilon$ = -0.22 × -0.35% ~ 0.08% is required. From the XRD measurements, we determined that the strain induced in the OOP direction of the NiFe lattice is: $(3.548 - 3.541)/3.548 \times 100 \sim 0.2\%$. Thus, the strain produced in the lattice is more than twice the amount needed to rotate the uniaxial anisotropy. Note that the IP strain is compressive and the OOP strain is tensile, consistent with the expected effects on deformations of isotropic materials. The use of an alloy specifically chosen for its optimal magnetic and mechanical properties, such as the Ni$_{90}$Fe$_{10}$ alloy, enables the effective utilization of BTO distortion to control its magnetic anisotropy. Fig. 3b confirms a strong magnetoelectric coupling between the layer and the substrate, sufficiently robust to rotate the IP uniaxial anisotropy by 90º. Importantly, these findings are reproducible, demonstrating that the direction of the uniaxial magnetic anisotropy can be effectively controlled through electrical methods, with the easy axis pointing along $\phi = 0º$ without poling and along $\phi = 90º$ when an electric field is applied.

Several hysteresis loops have been studied at different poling fields to investigate the VCMA. Fig. 3c shows the squareness ($M_{rem}/M_{sat}$) of the cycles obtained in a ferroelectric cycle starting from +0.8 MV/m along the $\phi = 0º$ direction. Low squareness values are associated with a hard axis, while high squareness values correspond to an easier direction. The graph confirms that the magnetic anisotropy of the sample is controlled using an electric field. Additionally, the lack of memory of the ferroelectric domains upon electric field removal allows the magnetic anisotropy to be tailored independently of the previous electric path taken, facilitating the achievement of the desired direction for the IP uniaxial anisotropy simply by either poling the heterostructure or not.

To calculate the converse magnetoelectric coupling, a 0.4 MV/m poling field is applied, which is sufficient to switch reproducibly from easy to hard axis (Fig. 3c), and we can thus calculate the converse magnetoelectric coupling [48] as:

$$\alpha_E = \frac{\mu_0 M_s \Delta S}{\Delta E} \tag{4}$$

Where $\Delta S = 0.7$ is the variation in squareness in the range $\Delta E = 0.4$ MV/m of the applied electric field needed to induce the uniaxial anisotropy. From equation 4 we obtain $\alpha_E = 1.43$ μs/m for the $Ni_{90}Fe_{10}$/BTO heterostructure, which exceeds values found in other BTO-based strain-coupled systems that do not contain rare-earth elements by one order of magnitude. For instance, the $\alpha_E$ values for reported epitaxial heterostructures such as Ni/BTO, $Fe_3O_4$/BTO, and $La_{0.67}Sr_{0.33}MnO_3$/BTO are 0.09 μs/m [49], 0.1 μs/m [50], and 0.23 μs/m [48]. The most direct comparison is with our previous result for electrodeposited $Ni_{90}Fe_{10}$ on BTO(011), which achieved 0.205 μs/m under the piezoelectric regime [39]. The main difference between the reported experiments and ours lies in the ferromagnetic layer, whose properties, such as magnetostriction and anisotropy, were carefully tailored by controlling its thickness, composition, and electrodeposition growth rate to maximize its magnetic response to the strain generated by the BTO. These results are particularly noteworthy because the thick, polycrystalline $Ni_{90}Fe_{10}$ layer makes it suitable for technology transfer, paving the way for its use in future VCMA applications.

**Conclusions**

This study reveals the magnetoelectric coupling in the $Ni_{90}Fe_{10}$/BTO(001) heterostructure and how it can be exploited to control uniaxial IP magnetic anisotropy through an electric field. The as-grown NiFe layer exhibits weak PMA that changes to uniaxial IP anisotropy when the BTO is poled. The BTO substrate, lacking polarization retention upon electric field removal, results in two distinct ferroelectric configurations: one with random ferroelectric domain orientation and another with all ferroelectric domains aligned in the electric field direction. These two ferroelectric configurations lead to two directions for the IP magnetic anisotropy in the NiFe layer that are 90º apart in the plane, *i.e.* IP uniaxial magnetic anisotropy. This 90º rotation of the IP anisotropy yields a converse magnetoelectric coupling of 1.43 µs/m, which is one order of magnitude higher than the values reported for other BTO strain-coupled heterostructures reported so far.

**Acknowledgements**

This work has been financially supported through the projects PID2021-122980OB-C51 (AEI/FEDER) and PID2021-122980OA-C54 (AEI/FEDER) of the Spanish Ministry of Science and Innovation. A. B. would like to acknowledge the funding received from the Ministry of Universities and the European Union-Next Generation for the Margarita Salas fellowship. These experiments were performed at BL24-CIRCE beamline at ALBA Synchrotron with the collaboration of ALBA staff. We thank the CAI X-ray Diffraction Unit and Ion Implantation Unit at Complutense University of Madrid for their valuable technical assistance.

**Conflict of Interest**

The authors have no conflicts to disclose.

**Author Contributions**

The manuscript was written through contributions of all authors. All authors have given approval to the final version of the manuscript.

A. Begué: Data Curation (lead); Analysis (lead); Investigation (lead); Methodology (lead); Writing – original draft (lead). M. W. Khaliq: Investigation; Methodology. N. Cotón: Investigation; Methodology. I. Lorenzo-Feijoo: Investigation; Methodology. M. A. Niño: Analysis; Investigation; Methodology; Funding acquisition; Writing – review & editing. M. Foerster: Analysis; Investigation; Methodology; Funding acquisition; Writing – review & editing. R. Ranchal: Data Curation; Analysis; Investigation; Methodology; Conceptualization (lead); Funding acquisition; Project administration (lead); Writing – review & editing.

**Data availability**. The data that support the findings of this study are available from the corresponding author upon reasonable request.

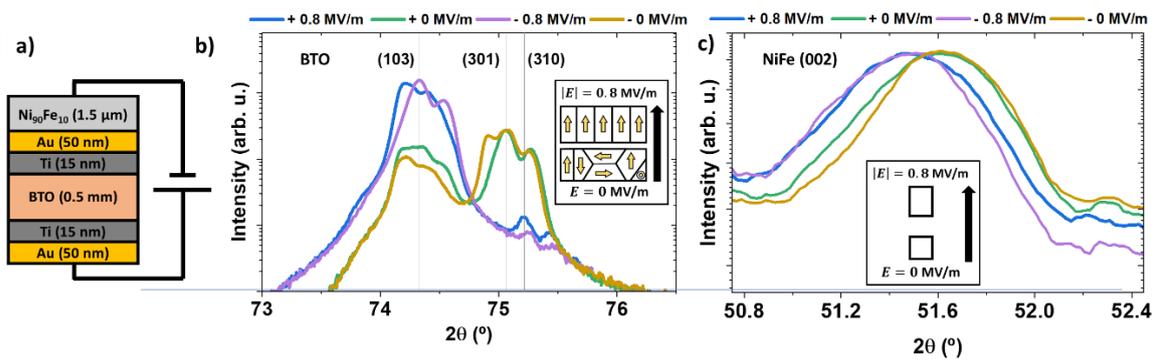

**Figure 1.** a) Schematic drawing of the studied Ni$_{90}$Fe$_{10}$/BaTiO$_3$(001) heterostructure, illustrating how the electric field is applied in the OOP direction. b) Reciprocal space XRD for BTO at several applied electric fields. Theoretical reflections {103} are shown to identify the peaks. Inset: Schematic cross-section of BTO lattice with poling (top) and without poling (bottom). Yellow arrows indicate the direction of the c-axis (polarization), and the black arrow indicates the direction of the applied electric field. c) XRD $\theta - 2\theta$ measurement of NiFe (002) reflection at several applied electric fields. Inset: Schematic cross-section of NiFe lattice with poling (top) and without poling (bottom). Black arrow indicates the direction of the applied electric field.

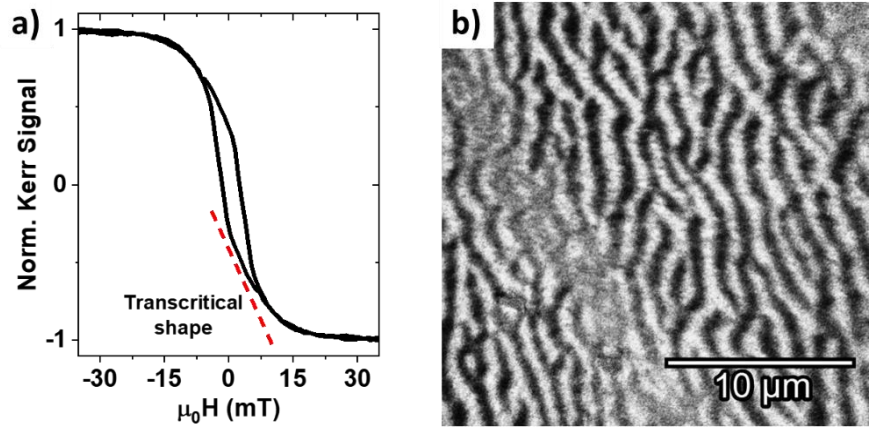

**Figure 2.** a) MOKE hysteresis loop for the as-grown $Ni_{90}Fe_{10}$ state, indicating the transcritical shape with a red dashed line. b) XMCD-PEEM image of $Ni_{90}Fe_{10}$ taken at magnetic remanence.

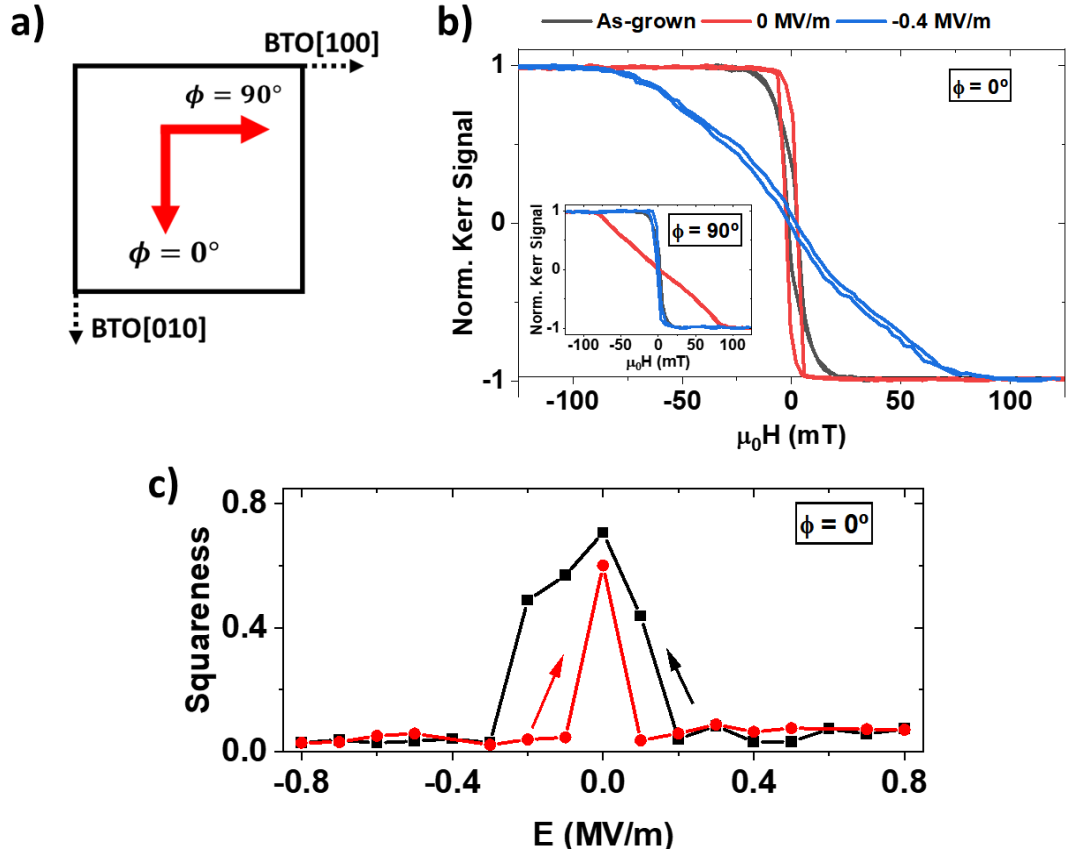

**Figure 3.** a) Schematic top view of the BTO substrate during saturated poling, indicating the IP directions corresponding to $\phi$ (red arrows) and the BTO crystallographic directions (dashed arrows). b) MOKE hysteresis loops of the as-grown state and the two states obtained after poling: without an applied electric field (0 MV/m) and with an electric field of -0.4 MV/m along $\phi = 0°$. Inset along $\phi = 90°$. The line colors in the main figure and inset correspond to the same legend. c) Squareness versus electric field obtained from several hysteresis loops ($\phi = 0°$). The cycle starts at 0.8 MV/m (black squares), and the return path from -0.8 MV/m (red dots) is indicated by arrows for clarity.